\documentclass[10pt]{iopart}

\usepackage[T1]{fontenc}
\usepackage[utf8]{inputenc}
\usepackage{graphicx}
\usepackage{tabularx}
\usepackage{color, soul}
\usepackage{textcomp}
\usepackage{enumerate}
\usepackage{todonotes}
\usepackage[draft]{hyperref}
\usepackage{comment}
\usepackage{subcaption}
\usepackage{url}
\usepackage[normalem]{ulem}
\usepackage{makecell}
\usepackage{graphbox}
\usepackage{stfloats}

\providecommand{\RV}[1]{#1}

\begin{document}

\title[Investigation of positive streamers in CO$_2$: experiments and 3D particle-in-cell simulations]{
Investigation of positive streamers in CO$_2$: experiments and 3D particle-in-cell simulations}

\author{Xiaoran Li$^{1, 2, 3}$, Siebe Dijcks$^{4}$, Anbang Sun$^{2}$, Sander Nijdam$^{4}$, Jannis Teunissen$^3$}

\address{
$^1$ Institute of Applied Physics and Computational Mathematics, Beijing 100094, People’s Republic of
China \\
$^2$ State Key Laboratory of Electrical Insulation and Power Equipment, School of Electrical Engineering, Xi'an Jiaotong University, Xi'an, 710049, China\\
	$^3$ Centrum Wiskunde \& Informatica, Amsterdam, The Netherlands\\
 $^4$ Eindhoven University of Technology, Eindhoven, The Netherlands}

\ead{anbang.sun@xjtu.edu.cn; jannis.teunissen@cwi.nl}

\vspace{10pt}
\begin{indented}
        \item[]\today
\end{indented}

\begin{abstract}
  We investigate the propagation of positive streamers in CO$_2$ through 3D particle-in-cell simulations, which are qualitatively compared against experimental results at 50\,mbar.
  The experiments show that CO$_2$ streamers are much more stochastic than air streamers at the same applied voltage, indicating that few electrons are available in front of the streamer head.
  In the simulations, we include a photoionization model for CO$_2$.
  The computational results show that even a small amount of photoionization can sustain positive streamer propagation, but this requires \RV{a background electric field close to the critical field. When we compare streamers in CO$_2$ and in air at the same applied voltage, the electric field at the streamer head and the electron density in the streamer channel are higher in CO$_2$.}
  We discuss the uncertainties in CO$_2$ photoionization and provide an estimate for the quenching pressure, which is based on the radiative lifetime of emitting states and the collision frequency of the gas.
  Furthermore, a criterion for self-sustained streamer growth in CO$_2$ is presented and compared against simulation results.
  
  
  


\end{abstract}

\ioptwocol

\section{Introduction}
\label{sec:introduction}


CO$_2$ is increasingly used as an insulating gas~\cite{seeger2016,seeger2015} in high-voltage equipment, and it is the main component of many alternatives to SF$_6$~\cite{Chachereau_2018,Zhang_2020}.
In several studies, electric breakdown properties of CO$_2$ have been measured at different pressures and temperatures and for different voltage waveforms~\cite{matsumura2005, okabe2007,goshima2008,hikita2008,wada2011,sun2015,kumar2021}.
We here focus on streamer discharges, which play an important role in the early stages of electric breakdown~\cite{nijdam2020}.



There are relatively few experimental studies on streamer discharges in CO$_2$.
A challenge is that such discharges are hard to image, due to their low light emission~\cite{kumar2021a}.
Seeger \textit{et al}~\cite{seeger2016} experimentally investigated the streamer stability field, streamer radius and velocity in 0.05-0.5\,MPa CO$_2$ at positive and negative polarity.
They found streamer stability fields of about 11 ± 2\,\mbox{V(m·Pa)$^{-1}$} for negative polarity.
For positive polarity, stability fields were up to a factor of two higher, depending on the pressure, which is in strong contrast with the behavior in air.
Mirpour \textit{et al}~\cite{mirpour2022} measured the delay in streamer inception in high-purity CO$_2$ at $0.3 \, \textrm{bar}$ for varying voltage waveforms, which included a `pre-pulse' before the main pulse.
The response to this pre-pulse was observed to be different in CO$_2$ than in air, which the authors relate to the different electron detachment mechanisms in these gases.



There are also few computational studies on streamers in CO$_2$.
Levko \textit{et al}~\cite{levko2017} investigated the branching of negative streamers in atmospheric-pressure CO$_2$, using 2D particle-in-cell simulations.
Photoionization was not included in these simulations, as it was argued to be negligible.
Bagheri \textit{et al}~\cite{bagheri2020a} simulated positive streamer propagation in atmospheric-pressure CO$_2$ with a 2D axisymmetric fluid model.
They also argued that there is negligible photoionization in gases with a large CO$_2$ fraction, and
therefore included different levels of background ionization (10$^9$ and 10$^{13}$\,m$^{-3}$).
With such background ionization, positive streamers were faster in CO$_2$ than in air when using the same background field, which seems to contradict the experimental findings of~\cite{seeger2016}.
However, the authors note that fluid simulations with low background ionization densities can be unrealistic~\cite{xiaoran-comparison}.
Recently, Marskar~\cite{Marskar_2023b} has performed impressive 3D simulations of streamers in CO$_2$ with a new ``kinetic Monte Carlo'' model, which approximates a fluid model with macroscopic particles.



\begin{figure}
	\centering
	\includegraphics[width=0.4\textwidth]{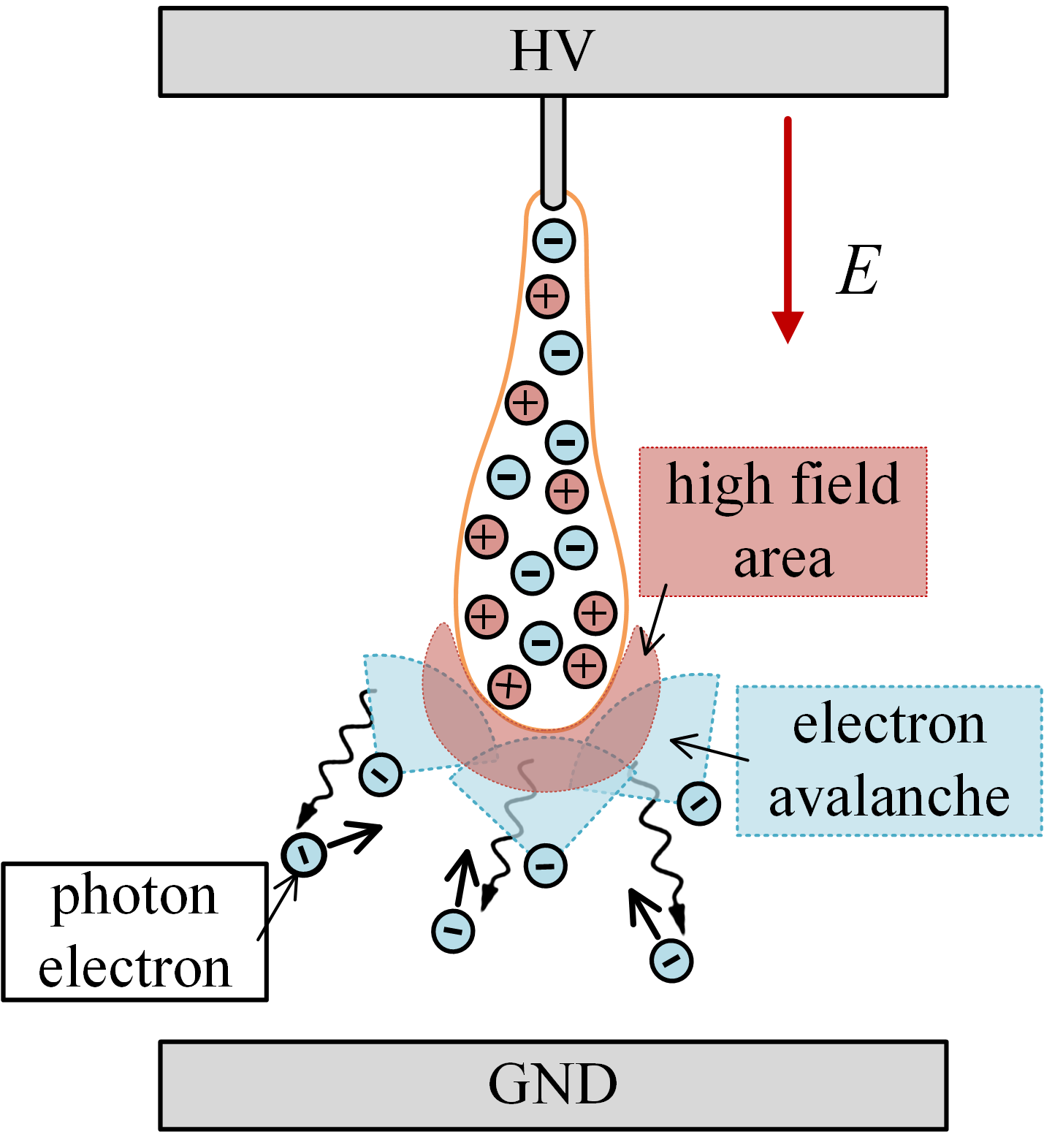}
	\caption{Schematic illustration of self-sustained positive streamer propagation due to photoionization.
          The streamer grows due to incoming electron avalanches, which should generate a sufficient number of photo-electrons ahead of the streamer so that the process is sustained.}
	\label{fig:mechanism}
\end{figure}

Positive streamers require a source of free electrons ahead of them to sustain their propagation~\cite{nijdam2020}, as illustrated in figure \ref{fig:mechanism}.
In air, such electrons are typically provided by photoionization, but in CO$_2$ photoionization is much weaker, since fewer ionizing photons are produced and because their typical absorption distance is orders of magnitude smaller~\cite{przybylski1962,pancheshnyi2014}.
In this paper, we focus on a particular question that was raised in~\cite{bagheri2020a}: can positive streamers propagate due to photoionization in CO$_2$?

To address this question, we simulate positive streamer propagation in CO$_2$ and we qualitatively compare against experimental measurements.
  Experimentally, a background electric field exceeding the critical field $E_\mathrm{cr}$ is used to enable single shot imaging of the faint (compared to air) streamers, at a pressure of 50\,mbar.
  A short camera gate time of 2.5\,ns allows us to resolve the shape of the streamer front and its propagation over time.
  Computationally, a conventional 3D particle-in-cell (PIC) model is used to study positive streamers under similar conditions as in the experiments, but in a smaller discharge gap due to the high computational cost.
  Photoionization is an important aspect of the model, but the parameters of photoionization in CO$_2$ are uncertain.
  We discuss these uncertainties, and investigate in particular the role of the quenching pressure.
  Furthermore, we present a criterion for self-sustained propagation due to photoionization.

	A difference between the present paper and~\cite{Marskar_2023b} is that we here use a conventional particle-in-cell model, whereas the kinetic Monte Carlo approach of~\cite{Marskar_2023b} uses particles to approximate the behavior of a fluid model.
    A conventional particle-in-cell model can capture the stochastic growth of discharges in CO$_2$ more accurately, but is computationally costly, so that we are limited to small-scale simulations.
    Furthermore, we provide a qualitative comparison with experimental results, and we provide additional discussion on quenching and self-sustained discharge growth.


  The paper is structured as follows.
  In section~\ref{sec:photoi_CO2}, two approaches for modeling photoionization in CO$_2$ are discussed as well as the available data.
  In section~\ref{sec:method}, the computational and experimental methodology are described.
  Simulations of positive streamers in CO$_2$ with different quenching pressures are presented in section~\ref{sec:results-quenching} and simulations in different background electric fields in section~\ref{sec:diff_E}.
  Furthermore, simulations in CO$_2$ and in air are compared in section~\ref{sec:air}, and experimental and computational results are compared in section~\ref{sec:exp-results}.
  We discuss the quenching pressure for photoionization in CO$_2$ in section~\ref{sec:quenching-discussion}, we briefly comment on background ionization in section~\ref{sec:background_ionization}, and we propose a self-sustaining criterion for streamer discharges in CO$_2$ in section~\ref{sec:criteria}.




\section{Photoionization in CO$_2$}
\label{sec:photoi_CO2}

\subsection{Description based on cross sections}
\label{sec:photoi-cross-sec}

To fully describe photoionization in CO$_2$, we in principle need to know quite a few things: the excited species that can emit ionizing photons, the cross-sections describing their generation, radiative lifetimes, quenching rates, and cross sections describing the absorption of these photons.
However, compared to air, there is relatively little information about photoionization in CO$_2$~\cite{pancheshnyi2014}.

The first ionization limit of CO$_2$ is 13.77\,eV, which corresponds to a photon wavelength of 90\,nm.
According to Pancheshnyi~\cite{pancheshnyi2014}, photons in the spectral range of 83-89\,nm can contribute to the photoionization of CO$_2$.
For these photons, the absorption coefficients range from 0.34 to 2.2 cm$^{-1}$Torr$^{-1}$, which corresponds to absorption lengths in the range of 6.1-40\,$\mu$m at standard temperature and pressure.
Note that in air at standard temperature and pressure, the corresponding absorption lengths are much longer, namely 33\,$\mu$m and 1.9\,mm.

According to the emission spectrum reported in ~\cite{sroka1970}, such ionizing photons can be produced by transitions of O$_\mathrm{II}$, C$_\mathrm{II}$ and O$_\mathrm{I}$ after the dissociation of a CO$_2$ molecule.
Dissociation cross sections contributing to VUV emission were reported in~\cite{sroka1970}.
These cross sections peak at energies of 100\,eV to 200\,eV, and they require a typical threshold energy of about 50\,eV.
It is however difficult to obtain accurate data near the lower threshold from the figures in~\cite{sroka1970}.

\subsection{Simplified photoionization model}
\label{sec:photoi-effective}

\begin{figure}
  \centering
  \includegraphics[width=0.5\textwidth]{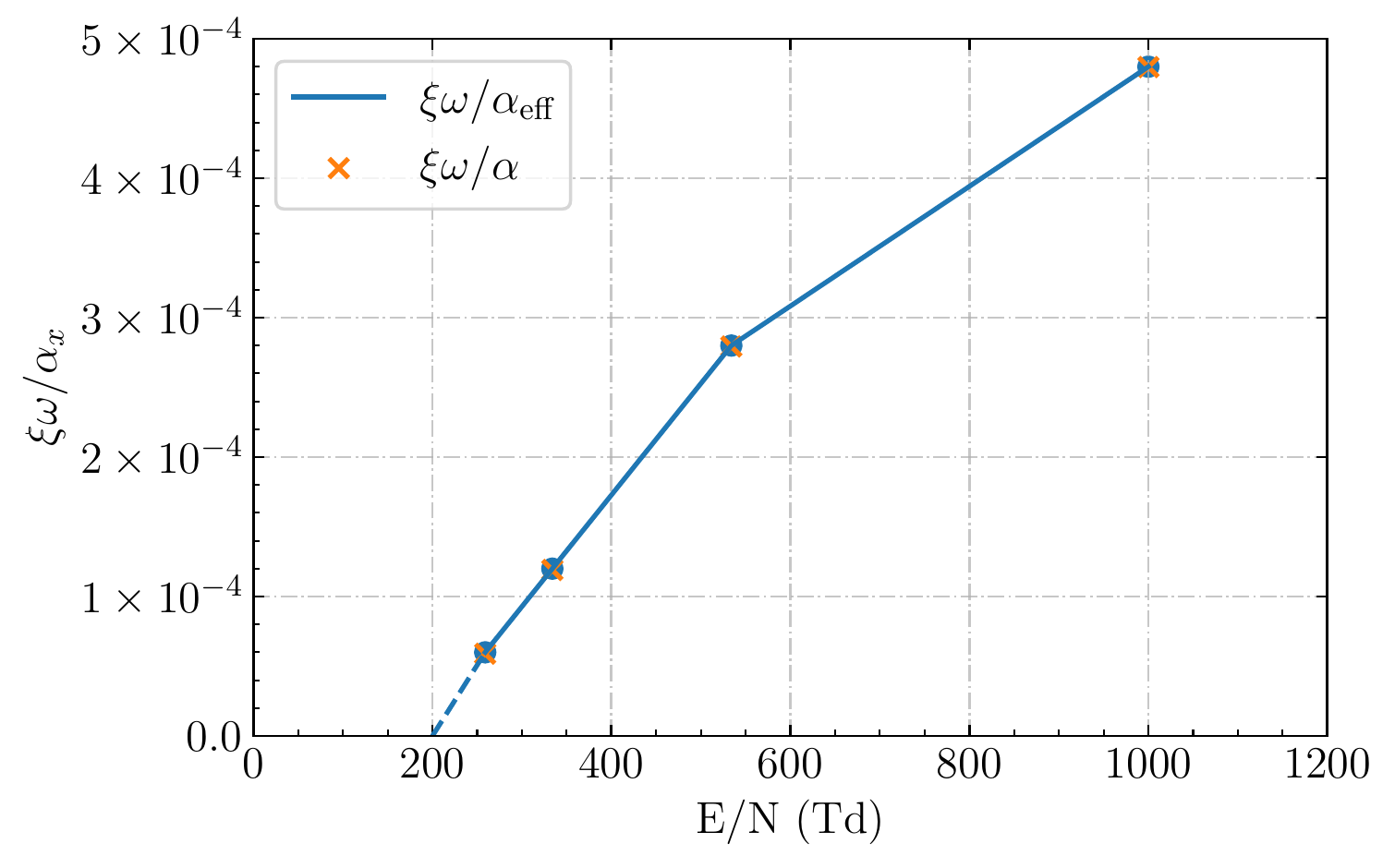}
  \caption{Coefficients $\xi\omega/\alpha_\mathrm{eff}$~\cite{przybylski1962,pancheshnyi2014} and $\xi\omega/\alpha$ (nearly identical), describing how many photoionization events are generated on average per (net) electron-impact ionization.
    We extrapolated $\xi\omega/\alpha_\mathrm{eff}$ to zero for $E/N = 200\,\mathrm{Td}$.}
  \label{fig:phoi_factor}
\end{figure}

Instead of a detailed description of all the involved processes and cross sections, a simplified description of photoionization can also be based on measurements of the amount of ionizing radiation in the vicinity of a discharge, like the Zheleznyak model for air~\cite{zhelezniak1982}.
Przybylski~\cite{przybylski1962} performed such measurements in CO$_2$ to obtain the ratio $\xi\omega/\alpha_\mathrm{eff}$ at pressures of about 1 to 3 Torr, with values between $0.6 \times 10^{-4}$ and $4.8 \times 10^{-4}$.
This ratio indicates how many photoionization events are generated (on average) per net electron-impact ionization, with $\alpha_\mathrm{eff}$ being the effective ionization coefficient, $\omega$ the production coefficient of photon-emitting states (with the same units as $\alpha_\mathrm{eff}$), and $\xi$ a numerical factor the probability of ionization after photon absorption.
The value of $\xi\omega/\alpha_\mathrm{eff}$ depends on the reduced electric field, and we follow the assumption made in~\cite{pancheshnyi2014} that the pressure range in~\cite{przybylski1962} corresponds to reduced electric fields between 260 and 1000\,Td, resulting in the field-dependence shown in figure~\ref{fig:phoi_factor}. Note that we extrapolated the ratio to zero for $E/N$ = 200\,Td, and that the ratio $\xi\omega/\alpha = \xi\omega/\alpha_\mathrm{eff} \cdot (\alpha_\mathrm{eff}/\alpha)$ is also shown.
In air, the value of $\xi\omega/\alpha_\mathrm{eff}$ ranges from 0.05 to 0.12 for $E/N$ between 20 and 600\,Td~\cite{zhelezniak1982}. The photoionization production in CO$_2$ is thus two to three orders of magnitude lower than in air.

\subsection{Discussion}
\label{sec:phoi-discussion}

For the simulations reported here, we use a photoionization model based on the data from~\cite{przybylski1962}.
It is difficult to assess the accuracy of the data in~\cite{przybylski1962} and~\cite{sroka1970}, since there appear to be no other sources to compare to.
We find that significantly more photons are generated with the parameters from~\cite{przybylski1962} than with the cross-sections from~\cite{sroka1970}.
A possible explanation could be that the cross sections in~\cite{sroka1970} are for CO$_2$ dissociation directly combined with the generation of an excited state (e.g., O$_\mathrm{II}$, C$_\mathrm{II}$).
In the experiments of Przybylski~\cite{przybylski1962}, a fraction of the CO$_2$ molecules might already have been dissociated, so that less energy would be required to produce the excited states. However, this can also be the case in our experiments.

There are several uncertainties when using data from either of these sources in a computational model.
First, there appears to be no information on the collisional quenching of emitting states in CO$_2$.
Second, for the approach based on cross sections~\cite{sroka1970}, there is uncertainty in the cross sections themselves (especially at low energies, as was already mentioned), and it is not clear whether the given cross sections provide a `complete' description of the process.
Third, for the measurements of~\cite{przybylski1962}, a factor $4\pi$ might have to be included in $\xi\omega/\alpha_\mathrm{eff}$, as discussed in~\cite{pancheshnyi2014}, and the purity of the CO$_2$ used in the experiments was not reported.
Furthermore, these results were obtained at different pressures and a constant current, and later converted into an $E/N$ dependence in~\cite{pancheshnyi2014}.
It is unclear how accurate this $E/N$ dependence is, in particular when used in a streamer discharge model in which fields vary rapidly in space and time.

\section{Methodology}
\label{sec:method}

\subsection{Experimental imaging of positive streamers in CO$_2$}
\label{sec:exp}

We generate positive streamers in CO$_2$ at 50\,mbar in a plate-plate geometry with a gap of 8\,cm between the plates and a 1\,cm long needle electrode protruding from the powered high-voltage plate.
A description of the electrode geometry, the discharge vessel, the voltage pulse generator and the imaging system can be found in~\cite{xiaoran-comparison, dijcks2023}.

A voltage pulse with an amplitude of 10\,kV is applied to the 8\,cm gap for 200\,ns with a repetition frequency of 20\,Hz.
The corresponding background electric field (average electric field between the plate electrodes) is $E_\mathrm{bg}$ of 1.25\,kV/cm, which is a bit higher than the critical electric field $E_\mathrm{cr}$ of CO$_2$ (1.1\,kV/cm at 50\,mbar~\cite{bagheri2020a}).
A background electric field above the critical field reduces inception jitter~\cite{seeger2016,mirpour2022} and increases the luminosity of the streamers, which are both essential for taking phase-resolved pictures.

The camera has a gate time of 2.5 ns and it is synchronized with the pulse generator so that one discharge is imaged per exposure.
After each pulse, the camera trigger delay is increased by 2.5 ns, so that a series of shots capture the time evolution of a discharge.
  We remark that streamers typically bridge the gap in about 50 ns, but the voltage pulse continues until 200 ns due to the limitations of our pulse generator.
  In the last 150 ns, the streamer paths transition into spark channels that locally heat the gas.

\subsection{PIC-MCC model}
\label{sec:pic-mcc-model}

We simulate positive streamer discharges in CO$_2$ using \texttt{afivo-pic}~\cite{teunissen2016,wang2022}, an open-source parallel AMR code for particle-in-cell discharge simulations with Monte-Carlo collisions (PIC-MCC).
We provide a brief summary of the model below, for a more detailed description see~\cite{teunissen2016,wang2022}.
Ions as tracked as densities, and they are assumed to be immobile on the nanosecond time scales considered in this paper.
Electrons are tracked as particles.
They are accelerated by the electric field and stochastically collide with a background of neutral gas molecules.
The electron coordinates and velocities are advanced with the `velocity Verlet' scheme, and isotropic scattering is assumed for collisions.
Electron-neutral scattering cross sections for CO$_2$ were taken from the IST-Lisbon database~\cite{grofulovic2016} on LXCat, \RV{which contains one ionization reaction ($\textrm{e} + \textrm{CO}_2 {\rightarrow} \textrm{e} + \textrm{e} + \textrm{CO}_2^+$ (13.80\,eV)), one dissociative attachment reaction ($\textrm{e} + \textrm{CO}_2 {\rightarrow} \textrm{CO} + \textrm{O}^-$ (3.90\,eV)) and thirteen excitation reactions.
In addition,} this database contains an effective momentum transfer cross section, accounting for the combined effect of elastic and inelastic processes~\cite{Pitchford_1982}.
We subtracted the sum of the inelastic cross sections to obtain an approximate elastic cross section that can be used in the particle model.


To make the particle-in-cell simulation computationally feasible, super-particles are used to represent several physical particles.
The weights of super-particles (how many physical particles they stand for) are adaptively controlled to obtain a desired number of simulated particles per cell, which was here set to 75.
Details of the weight control method can be found in~\cite{teunissen2016} and~\cite{wang2022}.
Adaptive mesh refinement is used for computational efficiency~\cite{teunissen2018}.
The mesh is refined if $\alpha(E)\Delta x > 0.8$, where $\alpha$ is the ionization coefficient.
This leads to a minimal grid size of around 10\,$\mu$m in our simulations.

\subsection{Photoionization model}
\label{sec:phoi_model}




In our simulations, we use a Monte Carlo photoionization procedure similar to the one described in~\cite{Chanrion_2008,teunissen2016}.
The photoionization model is based on the results of Przybylski~\cite{przybylski1962}, and we use the data for $\xi\omega/\alpha_\mathrm{eff}$ shown in figure~\ref{fig:phoi_factor}.
The model is implemented as follows.
If an electron super-particle with weight $w$ causes an impact ionization event, then the number of ionizing photons that are generated is sampled from a Poisson distribution with mean
\begin{equation}
  \label{eq:n-photons}
  N_\gamma = w \eta_q\xi\omega/\alpha.
\end{equation}
In this expression, the factor
\begin{equation}
  \label{eq:p-quenching}
  \eta_q = \frac{p_q}{p+p_q}
\end{equation}
accounts for collisional quenching, with $p_q$ the quenching pressure and $p$ the gas pressure.
Since the value of $p_q$ has not been determined by experiments, we investigate its effect in section~\ref{sec:results-quenching} by setting it to 50, 10, and 5\,mbar.
  Additionally, a rough estimation of $p_q$, based on the collision rate of gas molecules and the radiative lifetime of the emitting states, is made in Section~\ref{sec:quenching-discussion}.

Each produced photon starts at the location of the ionization, and its direction is sampled from an isotropic distribution.
Absorption coefficients $\mu$ are sampled as
\begin{equation}
  \mu = \mu_{\rm{min}} (\mu_{\rm{max}}/\mu_{\rm{min}})^{U_1},
\label{eq:absorption_coeff}
\end{equation}
where $U_1$ is a (0,1) uniform random number, and where $\mu_{\rm{max}}/p$ = 2.2\,cm$^{-1}$Torr$^{-1}$ and $\mu_{\rm{min}}/p$ = 0.34\,cm$^{-1}$Torr$^{-1}$ are the maximum and minimum absorption coefficients~\cite{chan1993,pancheshnyi2014}, as discussed in section~\ref{sec:photoi-cross-sec}.
Note that this sampling follows the linear approximation in figure 24 of~\cite{pancheshnyi2014}.
Afterwards, absorption distances are sampled from the exponential distribution as
\begin{equation}
  l = -\rm{ln}(1-U_2)/\mu,
  \label{eq:absorption_length}
\end{equation}
where $U_2$ is another (0,1) uniform random number.
Finally, a photoionization event is generated at the location where the photon is absorbed, generating a free electron and a positive ion.

\subsection{Simulation domain and initial conditions}
\label{sec:condition}

\begin{figure}[h]
	\centering
	\includegraphics[width=0.45\textwidth]{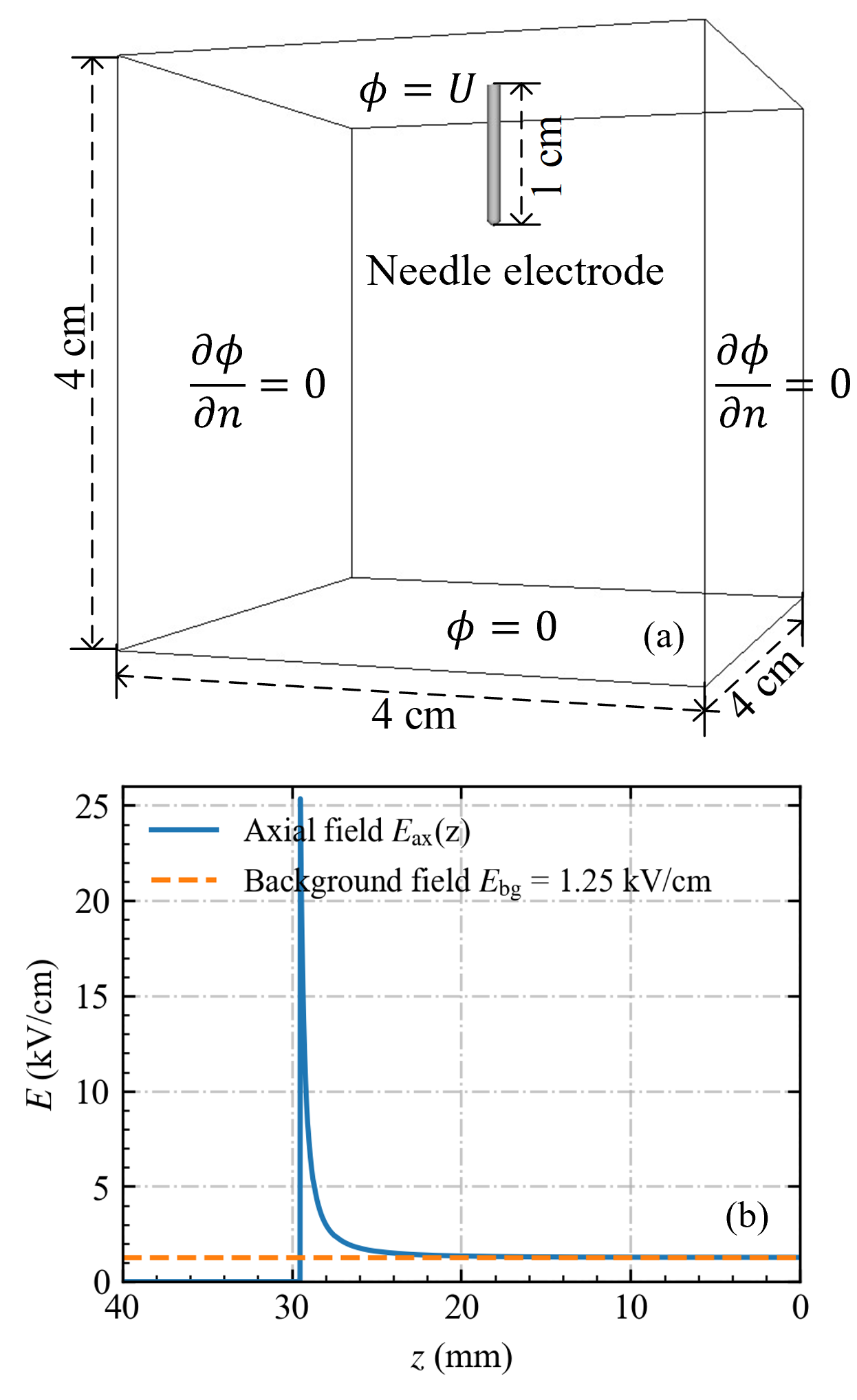}
	\caption{(a) The computational domain and boundary conditions for the electric potential. (b) The axial electric field profile $E_\mathrm{ax}$($z$) in domain with an applied voltage of 5\,kV. The corresponding background electric field $E_\mathrm{bg}$ is 1.25\,kV/cm, defined as the average electric field between the plate electrodes.}
	\label{fig:domain}
\end{figure}

We simulate positive streamers in CO$_2$ at 50\,mbar and 300\,K, using a $(4 \, \textrm{cm})^3$ 3D computational domain.
This domain and the boundary conditions for the electric potential are illustrated in figure~\ref{fig:domain}(a).
As in the experiments, we use a parallel-plate electrode geometry with a needle protrusion.
The 1\,cm long needle electrode is placed at the center of the powered electrode, with a 0.5\,mm radius and a rounded tip.

For reasons of computational efficiency, there are two main differences compared to the experimental geometry.
First, the gap size is smaller in the simulations, namely 4\,cm compared to 8\,cm in the experiments.
Second, the top and bottom plates in the simulations effectively have an infinite extent, due to the Neumann boundary conditions on the side of the domain, whereas the plate electrodes in the experiments have a radius of 4\,cm (powered electrode) and 6\,cm (grounded electrode).
The main computational limitation we faced was the memory cost of the simulations.
In a 4\,cm gap, the simulations in CO$_2$ already ran out of memory (64\,GB) when streamers approached the ground electrode, using up to 500 million particles.
To take into account the effect of finite plate electrodes in the experiments, the simulation domain would have to be about $(10 \, \textrm{cm})^2\times 8 \, \textrm{cm}$, requiring significantly more memory.




Figure~\ref{fig:domain}(b) shows the axial electric field profile $E_\mathrm{ax}$($z$) in our computational domain with an applied voltage of 5\,kV.
The corresponding background electric field (average electric field between the plate electrodes) $E_\mathrm{bg}$ is 1.25\,kV/cm, which is the same as the $E_\mathrm{bg}$ in the experiments.
The axial field converges to $E_\mathrm{bg}$ at places away from the needle tip.

As indicated in figure~\ref{fig:domain}(a), homogeneous Neumann boundary conditions are applied for the electric potential in the lateral directions.
Electrons are absorbed at electrodes, but not emitted.
To initiate the discharges, a neutral seed is used.
The seed consists of 5000 electron–CO$_2^+$ pairs at coordinates that are drawn from a Gaussian distribution centered at the tip of the needle electrode with a standard deviation of 2\,mm.
\RV{Additional simulations have shown that the initial seed has a slight effect on the lowest voltage required for the streamer to cross the gap, but it hardly impacts the streamer's later propagation, which is sustained by photoionization.}

\section{Results}
\label{sec:results}

\subsection{Simulations with different quenching pressures}
\label{sec:results-quenching}

\begin{figure*}[htb]
  \centering
  \includegraphics[width=1.0\textwidth]{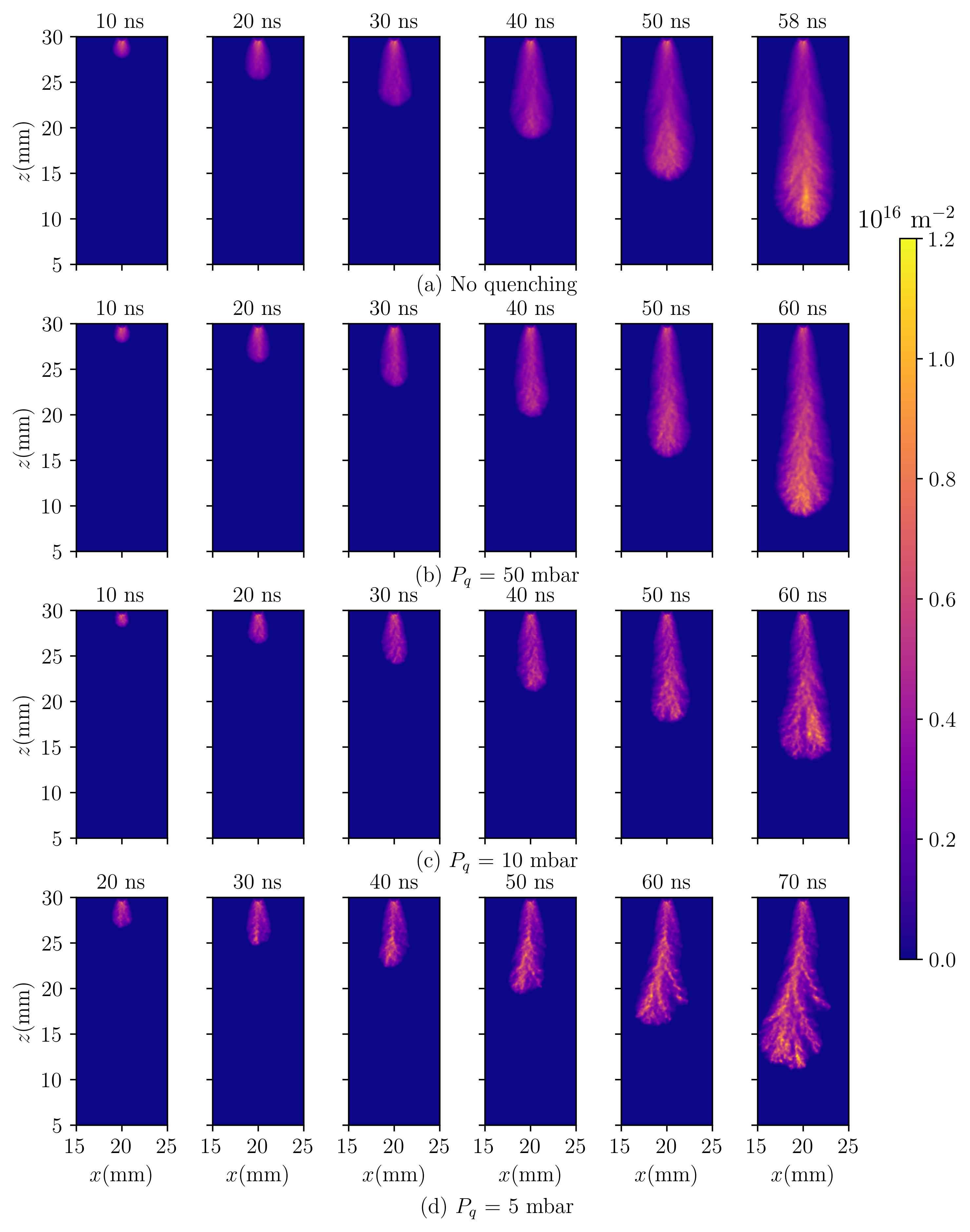}
  \caption{Time evolution of simulated positive streamers in CO$_2$ with quenching pressures of 50, 10 and 5\,mbar. Each sub-figure shows the electron density integrated along the $y$ direction in units of m$^{-2}$. Note that times are not always equal per column.}
  \label{fig:diff_pq}
\end{figure*}

We simulate streamers in pure CO$_2$ with an applied voltage of 5\,kV at 300\,K using the model and initial conditions described in section~\ref{sec:method}.
Since we do not know the quenching pressure $p_q$ and the value of the factor $\eta_q$ in equation~(\ref{eq:p-quenching}), we show results for several values: $p_q = 50$\,mbar ($\eta_q = 1/2$), $p_q = 10$\,mbar ($\eta_q = 1/6$) and $p_q=5$\,mbar ($\eta_q = 1/11$), as well as the case without quenching $p_q = \infty$ ($\eta_q = 1$).

The time evolution of streamers for these cases is shown in figure~\ref{fig:diff_pq}.
With a quenching pressure $p_q$ of 50\,mbar or no quenching, the streamer propagates downwards in a rather smooth and quasi-axisymmetry way.
As $p_q$ decreases, stochastic fluctuations in the electron density of the streamer channel increase.
For $p_q = 5$\,mbar, the symmetry of the streamer is broken and small branches can be observed when the streamer is longer than 15\,mm.
This increased stochasticity is caused by the reduction in the number of photoelectrons that are produced.
We suspect the streamers with a $p_q$ of 10 and 50\,mbar would also branch if they grow longer in a larger computational domain, but we could not verify this due to computational cost limitations.
To sustain the streamer's propagation with reduced photoionization, a higher electric field is required at the streamer head.
This results in a slightly higher electron density in the streamer channel.


Our results suggest that the velocity of positive streamers in CO$_2$ depends on the amount of photoionization, since the streamers with a lower quenching pressure are considerably slower.
Such a dependence is usually not observed in air, in which the velocity is typically not sensitive to the amount of photoionization, and in which increased photoionization can even reduce the streamer velocity, see e.g.~\cite{nijdam2010,xiaoran-comparison}.
This can be understood by considering two extremes.
If there are many photoelectrons, then the electron density ahead of a streamer will mostly grow due to electron impact ionization and it will not be sensitive to the number of photons.
But if there are almost no photoelectrons, then the local growth will often ``pause'' until a new photoelectron appears, with the duration of such pauses inversely proportional to the number of photons.
One could therefore expect the average streamer velocity to be inversely proportional to the number of photons, but this is too simplistic, since streamer properties and the properties of incoming electron avalanches also change depending on the number of photons.

\subsection{Effect of the background electric field}
\label{sec:diff_E}


We have studied the effect of the applied voltage by varying it in steps of 0.2\,kV in our
simulations.
At 3.4\,kV and without quenching, we observe the formation of a local ionized cloud around the needle tip which does not result in streamer propagation.
Such short positive discharges have also been observed in experiments~\cite{seeger2016}.
In this regime, the streamer is not self-sustained by photoionization, so there is only limited growth due to the electrons that are initially present.

Without quenching, the lowest voltage for a streamer to cross the gap is 3.6\,kV, which corresponds to a background field $E_{\rm{bg}} = 0.9 \, \mathrm{kV/cm}$.
With a quenching pressure of 5\,mbar, the lowest voltage for a streamer to cross the gap is 4\,kV, corresponding to a $E_{\rm{bg}}$ of 1.0\,kV/cm.
In both cases, the background electric field required for streamer crossing is close to the critical field $E_{\rm{cr}} = 1.1$\,kV/cm of CO$_2$ at 50\,mbar.
This is consistent with the experimental observation in~\cite{seeger2016} that positive streamers in CO$_2$ could only cross the gap when $E_{\rm{bg}}$ was close to $E_{\rm{cr}}$, using a quasi-uniform field setup.

\begin{figure*}[htb]
	\centering
	\includegraphics[width=1.0\textwidth]{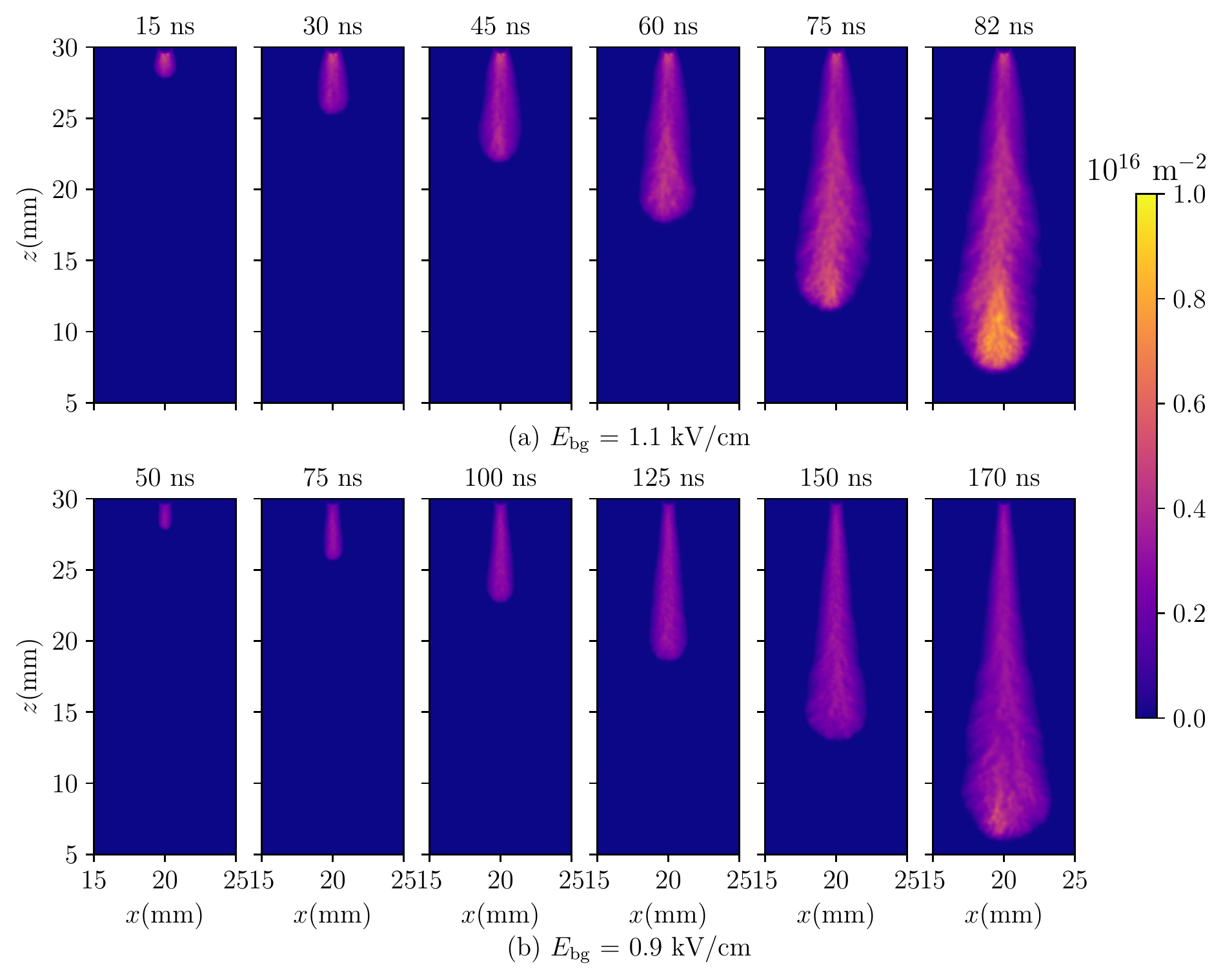}
	\caption{Time evolution of positive streamers in CO$_2$ in background electric fields of 0.9\,kV/cm and 1.1\,kV/cm, corresponding to applied voltages of 3.6\,kV and 4.4\,kV.
          The simulations were performed without quenching.
	Each sub-figure shows the electron density integrated along the $y$ direction.}
	\label{fig:diff_E}
\end{figure*}

Figure~\ref{fig:diff_E} shows the time evolution of positive streamers in CO$_2$ in a background electric field of 0.9 and 1.1\,kV/cm without quenching.
With a lower $E_{\rm{bg}}$, the streamer inception time increases significantly, the streamer is slower, and the streamer radius is initially smaller.

\RV{The voltage required to cross the gap slightly depends on the number of electrons in the initial seed.
  For the case without quenching, using 20,000 initial electron-CO$_2^+$ pairs (four times more than before) allows the streamer to cross the gap with an applied voltage of 3.4 kV, corresponding to $E_{\rm{bg}} = 0.85\,$kV/cm.
  The effect of the seed is small because it is localized near the electrode, so it does not contribute to the later streamer growth that has to be sustained by photo-electrons.
  Due to the weak photoionization in CO$_2$ and the short photon absorption length, this self-sustained growth seems to require a background field close to $E_{\rm{cr}}$.
  In contrast, the fluid simulations in~\cite{bagheri2020a} suggest that positive streamers can propagate in background fields well below $E_{\rm{cr}}$ when a background density of electrons is present homogeneously in the domain.
  Experimental measurements of the minimal streamer crossing field could therefore provide information on the source of free electrons under different conditions: crossing fields well below $E_{\rm{cr}}$ would indicate that another mechanism (such as detachment) is providing free electrons instead of photoionization.
}


\subsection{Comparison with streamer simulations in air}
\label{sec:air}



\begin{figure*}[htb]
	\centering
	\includegraphics[width=1.0\textwidth]{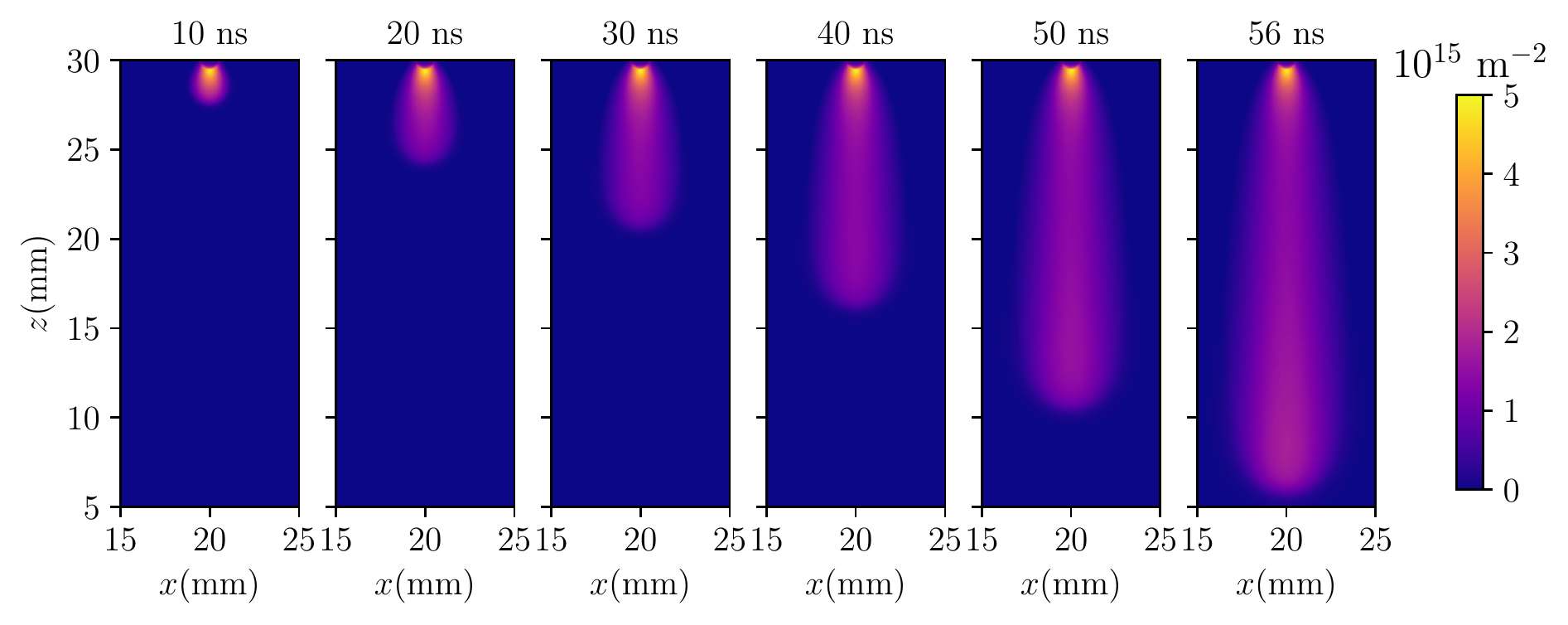}
	\caption{The evolution of streamers in air at the same pressure and applied voltage as for the CO$_2$ streamers in figure~\ref{fig:diff_pq}.
          Each sub-figure shows the electron density integrated along the $y$ direction.}
	\label{fig:sim_air}
\end{figure*}

In this section, we simulate a positive streamer in air under the same conditions (voltage, gas pressure) as for the CO$_2$ streamers in section~\ref{sec:condition}.
Photoionization is modeled with the Zheleznyak model for air~\cite{zhelezniak1982} using a Monte Carlo approach~\cite{teunissen2016} and with a quenching pressure of 40\,mbar.
Phelps' cross sections for N$_2$ and O$_2$~\cite{Phelps-database} are used for electron-neutral collisions.
The resulting streamer evolution in air is shown in figure~\ref{fig:sim_air}.
Compared with the CO$_2$ streamer without quenching in figure~\ref{fig:diff_pq}(a), the streamer velocity in air is only slightly higher.
The main difference is that the streamer propagation in air is much smoother than in CO$_2$.
It almost has a fully axisymmetric shape and hardly any density fluctuations are observed.


\begin{figure*}[h]
	\centering
	
	\includegraphics[width=1.0\textwidth]{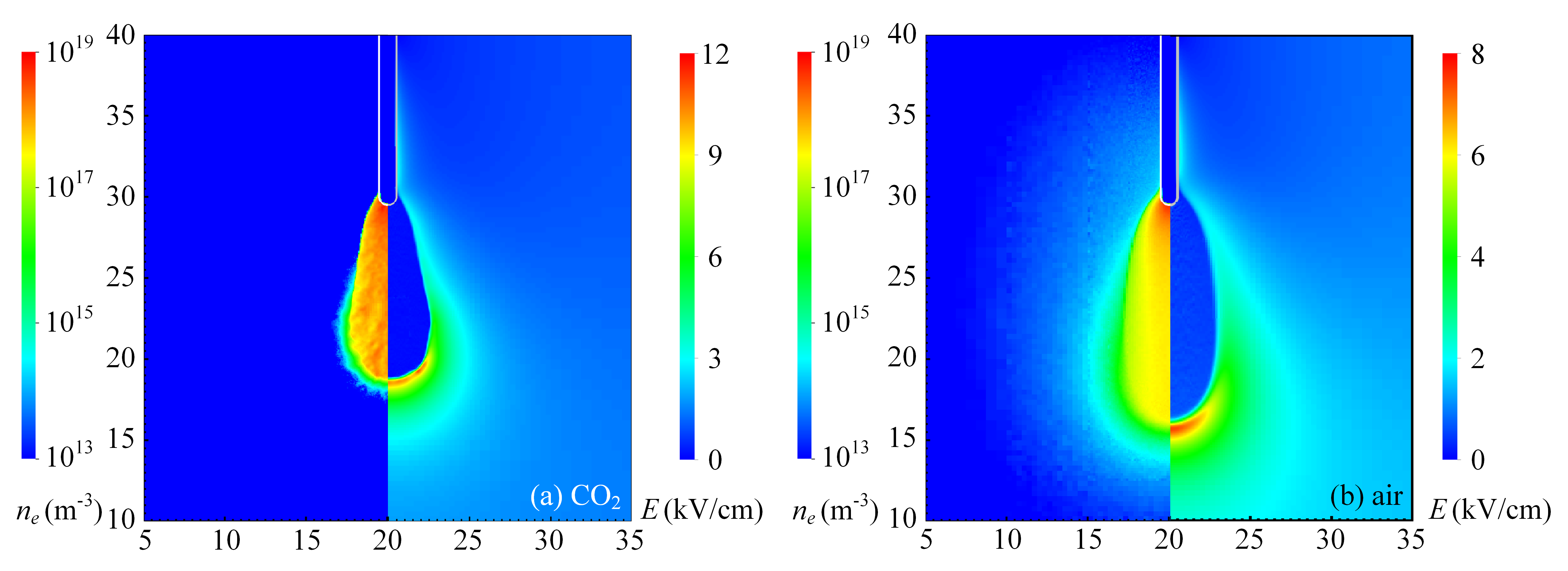}
	\caption{Electric field and electron density profiles of streamers in (a) CO$_2$ and (b) air at 40\,ns, sliced at y = 20\,mm. The streamers in CO$_2$ and air correspond to figure~\ref{fig:diff_pq}(a) and figure~\ref{fig:sim_air}, respectively.}
	\label{fig:CO2_air_ne_E}
\end{figure*}

We compare the electric field and the electron density profile of streamers in CO$_2$ and in air at 40\,ns in figure~\ref{fig:CO2_air_ne_E}.
The electron density in the CO$_2$ streamer channel is around $2 \times 10^{18}$\,m$^{-3}$ and the maximum electric field at the streamer head is around 10-12\,kV/cm during the entire evolution, which is much higher than the electron density ($3\times 10^{17}$\,m$^{-3}$) and maximum electric field (8\,kV/cm) of the positive streamer in air under the same conditions.
The electron density gradient around the streamer edge in CO$_2$ is steeper than in air, as was also observed to a lesser extent in simulations in air-methane mixtures~\cite{bouwman2022}.
This happens because photoionization is much weaker in CO$_2$ than it is in air, mostly due to the short absorption distance of ionizing photons.
Since relatively few photons are absorbed at a sufficient distance to contribute to discharge growth, the streamer head in CO$_2$ is less smooth than in air, and it ``wiggles'' forward.

\subsection{Qualitative comparison with experiments}
\label{sec:exp-results}

\begin{figure*}[htb]
	\centering
	\includegraphics[width=1.0\textwidth]{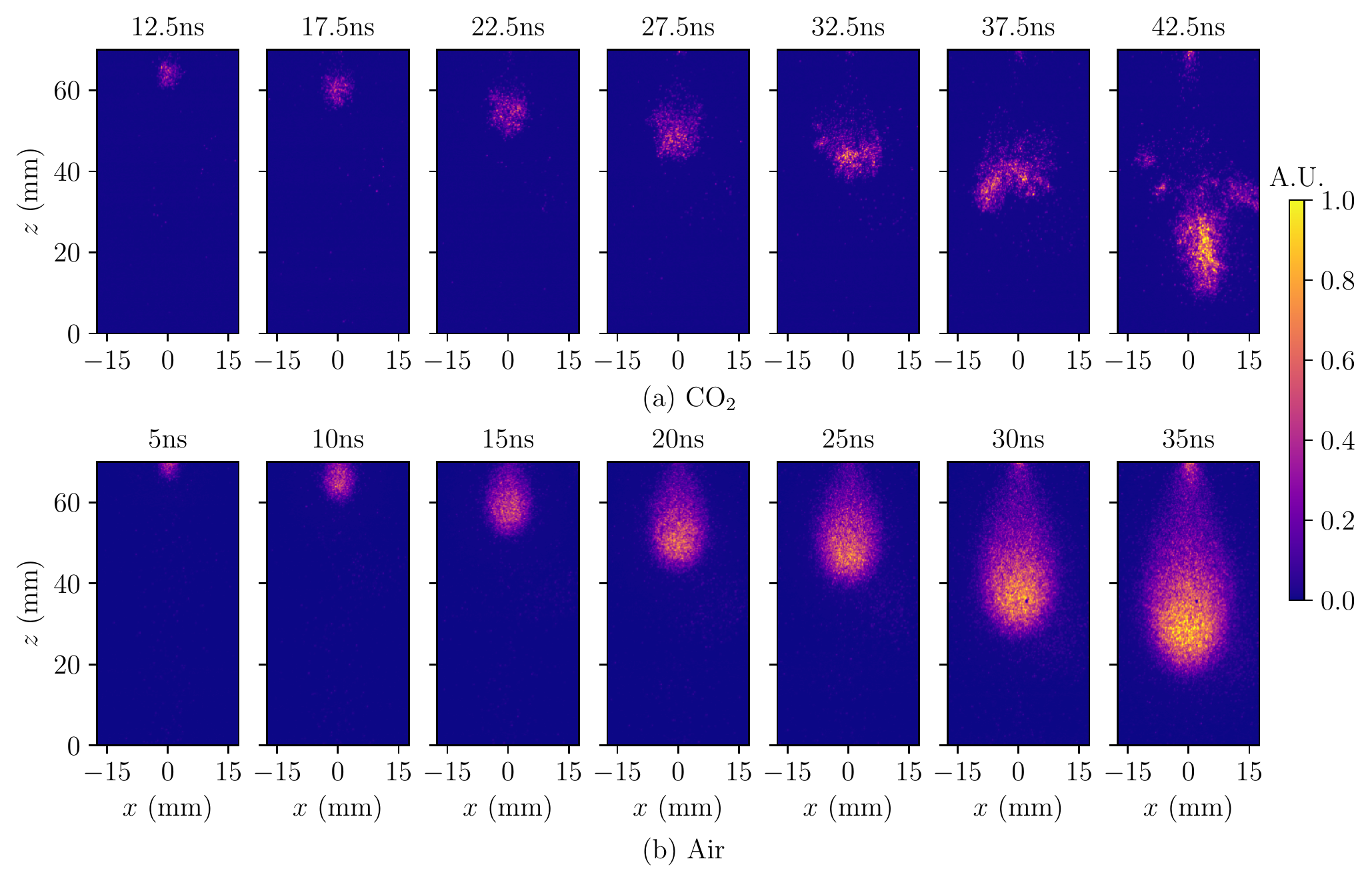}
	\caption{Experimental images of positive streamers in CO$_2$ (a) and in air (b) at 50\,mbar and $10 \, \textrm{kV}$, with a camera gate time of 2.5\,ns. The light emission in different subfigures is normalized per row to arbitrary units.
          In these images, $t = 0$ corresponds to the first light emission near the needle electrode.
          Several pulses were already applied before these images were captured.
        }
	\label{fig:exp_figure}
\end{figure*}

Experimental images of CO$_2$ streamers are shown in figure~\ref{fig:exp_figure}.
For comparison, streamers in air under the same conditions (voltage, pressure) are also shown.
We have normalized the light emission \textit{per row} to arbitrary units, so that frame-to-frame brightness variations are conserved, but the figure does not show that the air streamers are actually much brighter in the UV-vis range.

From the images, we can observe the development of the streamers over time.
In CO$_2$, this development is quite stochastic.
The streamer initially has one main channel, but later on it breaks into several branches, especially after 27.5\,ns.
In contrast, the air streamer smoothly grows downwards while its radius increases.
This difference indicates that fewer free electrons are available ahead of a positive CO$_2$ streamer, as discussed in section~\ref{sec:photoi_CO2}.

For both streamers in air and in CO$_2$, the streamer velocity and radius increase as the streamers grow.
During the first 20\,mm of propagation shown in figure~\ref{fig:exp_figure}, the average streamer is around 1$\times$10$^6$\,m/s for both CO$_2$ and air, and the streamer diameters increase from about 5\,mm to 15\,mm.
We can qualitatively compare these values with the simulation results presented in sections~\ref{sec:results-quenching} and~\ref{sec:air}.
In the simulations, average streamer velocities during the first 20\,mm of propagation are about $4 \times 10^5 \, \textrm{m/s}$ in CO$_2$ and $5 \times 10^5 \, \textrm{m/s}$ in air, and streamer diameters increase from about 2\,mm to 6\,mm in both gases.
Both the velocity and the radius are thus a factor of 2 to 2.5 lower in the simulations.

The main reason for this difference is probably the different geometry used in the simulations, in which an approximately homogeneous field is present between the plate electrodes (locally enhanced by the needle).
In the experiments, the powered plate electrode has a radius of 4\,cm, which means that the background electric field close to this plate will be higher.
Streamers will therefore be faster and wider with such a finite plate electrode.
In a previous validation study using the same experimental geometry, we have demonstrated that the velocity and radius can be a factor two larger due to this effect, see section 4.8 of~\cite{xiaoran-comparison}.
\RV{Another difference is that the experiments utilized 20\,Hz repetitive pulses combined with a 200\,ns pulse duration. During each pulse, the streamer crossed the gap in 50\,ns, subsequently transitioning to a spark discharge that heated the gas. From the voltage–current waveform, we estimate that about 15\,mJ was deposited in the plasma per pulse. At 20 Hz repetition frequency, this corresponds to $P$ = 0.3\,W of heating power. This 0.3\,W can lead to a temperature increase of approximately $70 \, \mathrm{K}$, based on a rough estimation as conducted in~\cite{xiaoran-comparison}. Such an increase can also contribute to the experimentally larger streamer velocity and diameter, as detailed for air in~\cite{xiaoran-comparison}.
Additionally, the accumulation of leftover species from repetitive pulses may further explain the observed discrepancies.}

We can also qualitatively compare the streamer morphology in CO$_2$ between simulations and experiments.
The rather stochastic development in the experiments seems to be more consistent with the simulations at relatively low quenching pressures (5\,mbar and 10\,mbar) than those with less quenching.
This could suggest that either quenching is very effective in CO$_2$ (because the quenching pressure is much lower than in air), or that the production of photoionization should be lower than in our model, meaning that figure~\ref{fig:phoi_factor} is an over-estimate.
However, the increased branching in the experiments might also be related to the different electrode geometry that was used, so further work is needed to better estimate $p_q$.

\section{Discussion}
\label{sec:discussion}

\subsection{An expression for the quenching pressure}
\label{sec:quenching-discussion}

Excited states that can emit ionizing photons can de-excite due to collisions with gas molecules.
Such quenching is often described with a single parameter, the quenching pressure $p_q$, using equation~(\ref{eq:p-quenching}).
For air, a quenching pressure $p_q$ of 40\,mbar is typically used~\cite{zhelezniak1982}.
However, as was mentioned in section~\ref{sec:photoi_CO2}, there appear to be no measurements of $p_q$ in CO$_2$.




If we know the frequency of collisions between gas molecules as well as the radiative lifetime of an excited state, we can obtain a rough estimate for the quenching pressure.
There are different ways to define a mean free path for gas molecules.
One way is to assume hard-sphere collisions and to then determine the mean free path that results in a viscosity consistent with measurements.
For air, this results in $l_\mathrm{air} \approx 68 \, \mathrm{nm}$ at atmospheric pressure and room temperature, and for CO$_2$, the result is $l_{\mathrm{CO}_2} \approx 45 \, \mathrm{nm}$.
The corresponding collision frequency of molecule-molecule collisions is given by $\nu_c = \sqrt{2} v/l$, where $v = \sqrt{\frac{8 k_B T}{\pi m}}$ is the average velocity of molecules with mass $m$.
For air under the given conditions, $v \approx 4.7 \times 10^2 \, \mathrm{m/s}$ while for CO$_2$ we have $v \approx 3.8 \times 10^2 \, \mathrm{m/s}$.
This results in collision frequencies $\nu_{c,\mathrm{air}} \approx 1.0 \times 10^{10} \, \mathrm{s}^{-1}$ and $\nu_{c,\mathrm{CO}_2} \approx 1.2 \times 10^{10} \, \mathrm{s}^{-1}$, and collision times $\tau_c$ given by their respective inverses.
In air, the radiative lifetime of the excited N$_2$ state emitting ionizing photons is about $\tau_\mathrm{rad} \approx 0.9 \, \textrm{ns}$~\cite{jiang2018, stephens2016}.
If the probability of quenching per molecule-molecule collision is $f$, then the effective collision time for quenching is $\tau_c' = \tau_c/f$.
This means that the probability of emitting a photon (without collisional quenching) is given by $\tau_c' / (\tau_\mathrm{rad} + \tau_c')$, which by definition should be equal to the factor $p_q/(p+p_q)$ from equation (\ref{eq:p-quenching}).
We can solve for $f$ to obtain
\begin{equation}
  f = \frac{\tau_c}{\tau_\mathrm{rad}} \frac{p}{p_q},
  \label{eq:quench-approx-f}
\end{equation}
which for air gives $f = 2.8$.
That we obtain $f > 1$ indicates that it is possible for molecules to de-excite other molecules at interaction distances at which no significant scattering takes place.


We can now use the above ideas to obtain a rough estimate for quenching in CO$_2$.
For the dissociated fragments of CO$_2$, radiative lifetimes were reported in~\cite{lawrence1970, heroux1969}:
1--3\,ns for O$_\mathrm{I}$(88nm), $1 \, \textrm{ns}$ for O$_\mathrm{II}$(83.3nm) and $0.3 \, \textrm{ns}$ for C$_\mathrm{II}$(90.4nm).
We do not know the respective contributions of these states to the photoionization in CO$_2$, but it seems reasonable to assume $\tau_\mathrm{rad} \sim 1 \, \textrm{ns}$.
We can then obtain an estimate for the quenching pressure by rewriting equation~(\ref{eq:quench-approx-f}) as
\begin{equation}
  p_q = \frac{p}{f} \frac{\tau_c}{\tau_\mathrm{rad}}.
  \label{eq:quench-approx-f}
\end{equation}
Using the same value of $f = 2.8$ as for air, this results in $p_q \sim 3 \times 10^{-2} \, \textrm{bar}$.
There is of course considerable uncertainty in this estimate, since the actual values of $f$ and $\tau_\mathrm{rad}$ are unknown.

\subsection{Background ionization in CO$_2$}
\label{sec:background_ionization}


Background ionization levels in ambient air (by radioactivity and cosmic rays) were reported to be 10$^9$ - 10$^{10}$\,m$^{-3}$ at ground level~\cite{pancheshnyi2005a}.
The background ionization level in an experimental vessel filled with CO$_2$ from a high-pressure gas cylinder is probably significantly lower, as the gas would contain less radon than ambient air.
Experimentally, large statistical time lags have been observed for the inception of positive streamers in CO$_2$~\cite{seeger2016,mirpour2022}, which suggests that there are typically few free electrons.

For positive streamer propagation, the composition of the background ionization also matters.
In electro-negative gases such as air, background ionization is mostly present in the form of positive and negative ions.
However, in pure CO$_2$, electron attachment is always dissociative \mbox{CO$_2$ + e $\rightarrow$ CO + O$^-$~\cite{chantry1972}}, which means that it only occurs for electrons with energies above 3 to 4\,eV.
In the absence of an electric field, we therefore expect the main electron loss processes to be wall losses and electron-ion recombination.
If there has previously been a discharge, some of the CO$_2$ will have dissociated, so that electron attachment to oxygen can also be important.

	
There is considerable uncertainty in all the above processes, making it difficult to provide generic estimates for background ionization levels in CO$_2$. Therefore, we focus on the effects of photoionization in this paper and assume negligible background ionization.


\subsection{A self-sustaining criterion for streamer discharges in CO$_2$}
\label{sec:criteria}

Self-sustained positive streamer growth requires $N_\mathrm{sp}\geq~1$, where $N_\mathrm{sp}$ is the number of new photoionization events that will on average be produced due to a single initial photoionization event.
We present an approximate criterion for such self-sustained growth based on the work of Naidis~\cite{naidis2005}, in which criteria for discharge inception in air near spherical and cylindrical electrodes were presented.
Our criterion has the following form:
\begin{equation}
\label{eq:growth-criterion}
N_\mathrm{sp} = f_\mathrm{g} \, \eta_q \xi\omega/\alpha_\mathrm{eff}
\int_{0}^{r_c} p(r) N_e(r) \mathrm{d}r \geq 1,
\end{equation}
where $r = 0$ corresponds to the streamer head position (the location of its charge layer), $r_c$ is the distance at which the electric field equals the critical field, $p(r)$ is the probability of photon absorption at a distance $r$ and $N_e(r)$ is the final size of an electron avalanche starting at location $r$ traveling to $r = 0$.
Furthermore, $f_\mathrm{g}$ is a geometric factor, $\eta_q$ the quenching factor and $\xi\omega/\alpha_\mathrm{eff}$ is the photoionization factor given in figure~\ref{fig:phoi_factor}.
The idea underlying equation (\ref{eq:growth-criterion}) is that a photoionization event produces an electron avalanche growing towards the streamer head, and the expected number of photons produced by this avalanche is computed.

Equation~(\ref{eq:growth-criterion}) is a simplified version of an equation given in~\cite{naidis2005}.
As in~\cite{naidis2005}, we assume that new photons are produced at $r = 0$, since the number of electrons grows exponentially in an avalanche.
The factor $\xi\omega/\alpha_\mathrm{eff}$ therefore depends on the field at the streamer head.
The probability of photon absorption is given by~\cite{zhelezniak1982}
\begin{equation}
  \label{eq:p-absorption}
  p(r) = \frac{\exp(-\mu_\mathrm{min} p r) - \exp(-\mu_\mathrm{max} p r)}
  {r \log(\mu_\mathrm{max}/\mu_\mathrm{min})},
\end{equation}
where $\mu_\mathrm{min}$ and $\mu_\mathrm{max}$ are the pressure-reduced absorption coefficients given in section \ref{sec:phoi_model}.

A simplification made here is the geometric factor $f_\mathrm{g}$.
Since streamers do not have the spherical or cylindrical geometry assumed in~\cite{naidis2005}, we use a simple line integral along their axial direction.
We roughly correct for geometric effects by assuming only photons within an angle $\theta_\mathrm{max} = 20^\circ$ contribute, so that $f_\mathrm{g} = \sin(\theta_\mathrm{max}/2)^2 \approx 0.03$.

Another modification is that we limit the size of avalanches to $N_\mathrm{max}$, so that
\begin{equation}
  \label{eq:avalanche-size}
  N_e(r) = \min\left[N_\mathrm{max},\,
    \exp\left( \int_0^{r} \alpha_\mathrm{eff}(r') \mathrm{d}r' \right)\right].
\end{equation}
As mentioned in~\cite{naidis2005}, the exponential growth of avalanches stops when space charge effects become important, see e.g.~\cite{Montijn_2006a}.
The number of electrons in a streamer will depend on its length, radius and degree of ionization, but for simplicity, we use a single value $N_\mathrm{max} = 10^8$.
Without such a limitation, there would be contributions of unrealistically large avalanches to equation~(\ref{eq:growth-criterion}), due to the strong field around a streamer.
Note that equation~(\ref{eq:avalanche-size}) is equivalent to limiting the $\alpha$-integral to a value $\log(N_\mathrm{max}) \approx 18.4$.

\begin{figure}
	\centering
	\includegraphics[width=0.5\textwidth]{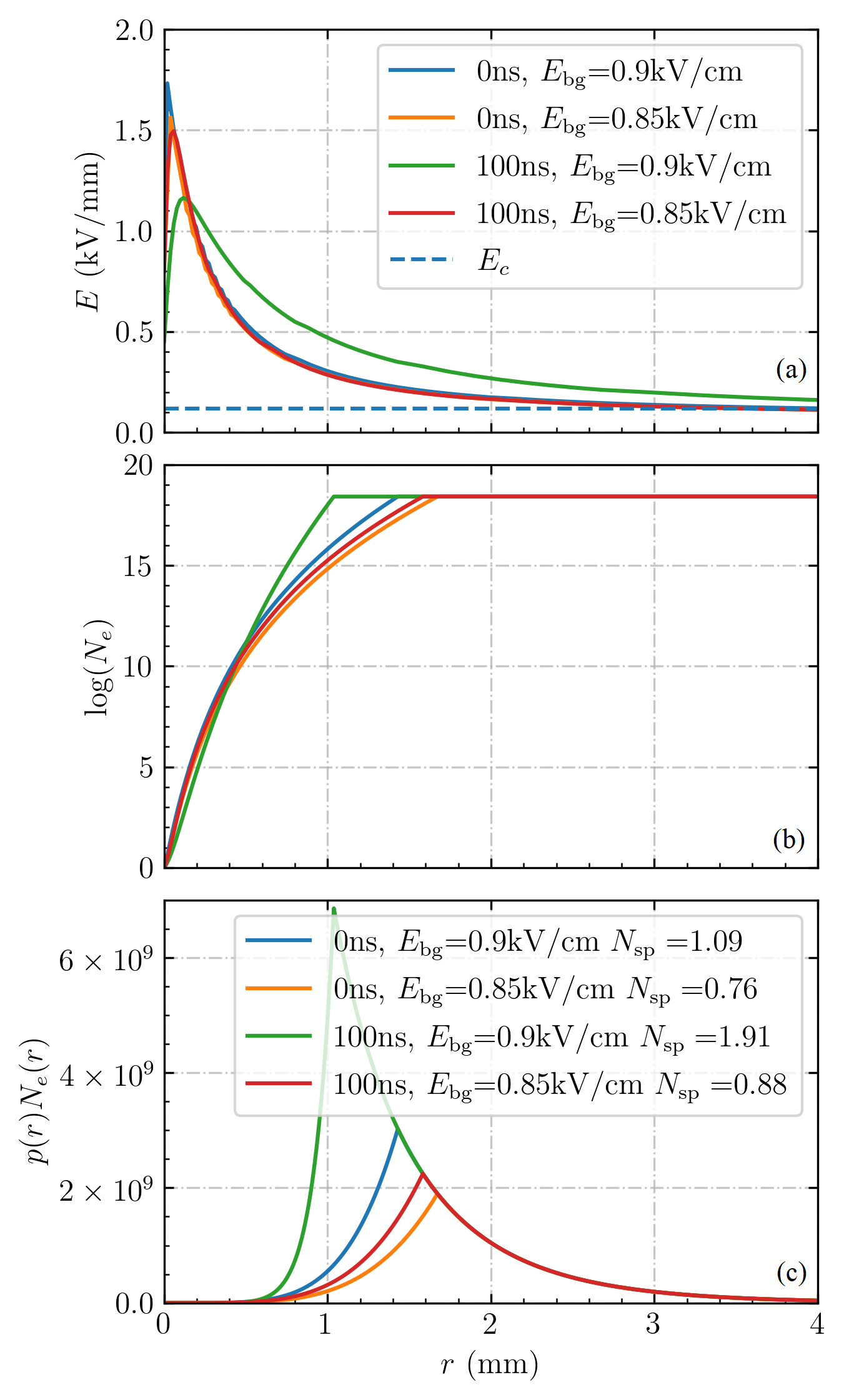}
	\caption{(a) Electric field strength versus vertical distance from the streamer head, extracted from the simulations with an $E_\mathrm{bg}$ of 0.9\,kV/cm and 0.85\,kV/cm at 0\,ns and 100\,ns.
          (b) Logarithm of electron avalanche size $N_e(r)$ according to equation (\ref{eq:avalanche-size}) versus the start position of the avalanche.
          (c) The integrand $p(r) N_e(r)$ of equation~(\ref{eq:growth-criterion}), where $p(r)$ denotes the probability of photon absorption at a distance $r$.}
	\label{fig:criterion}
\end{figure}

Figure~\ref{fig:criterion} illustrates the criterion of equation~(\ref{eq:growth-criterion}).
For several electric field profiles, the value of $\log(N_e)$ and the integrand $p(r) N_e(r)$ of equation~(\ref{eq:growth-criterion}) are shown, and the resulting values for $N_\mathrm{sp}$ are given.
The electric field profiles were extracted from the simulations with $E_\mathrm{bg} = 0.9\,\mathrm{kV/cm}$ and $E_\mathrm{bg} = 0.85\,\mathrm{kV/cm}$ at 0\,ns and 100\,ns, starting from the streamer head.
The values of $N_\mathrm{sp}$ suggest that the streamer is initially barely self-sustained ($N_\mathrm{sp}\sim1$).
At $t = 100 \, \textrm{ns}$, we find that ${N_\mathrm{sp}>1}$ for the 0.9\,kV/cm case, whereas the streamer cannot sustain itself ($N_\mathrm{sp} < 1$) for the 0.85\,kV/cm case.
This agrees with our simulation results for the latter case, in which a local ionized cloud formed around the needle tip without further streamer propagation.
Note that there is some uncertainty in equation~(\ref{eq:growth-criterion}), mostly due to the assumptions about the geometric factor $f_\mathrm{g}$ and due to the simple assumption of a maximum avalanche size $N_\mathrm{max}$, but it agrees well with simulation results with the present parameters.



\section{Conclusions}
\label{sec:conclusion}

We have performed 3D PIC-MCC simulations of positive streamer discharges in CO$_2$ at 50\,mbar, which we qualitatively compared against experiments.
In the particle model, photoionization was included based on the measurements of Przybylski~\cite{przybylski1962} and Teich~\cite{teich1967}, as interpreted by Pancheshnyi~\cite{pancheshnyi2014}.
An uncertain parameter is the quenching pressure $p_q$ for photoionization, for which we considered several values.
The simulations and experiments were both performed in a background field of $1.25 \, \textrm{kV/cm}$, which is above the critical field $E_\mathrm{cr} = 1.1 \, \mathrm{kV/cm}$ of CO$_2$ at 50\,mbar.
However, the experiments were performed in a bigger gap (8\,cm vs 4\,cm), and also with stronger field enhancement near the powered electrode.
The scale of the simulations was mostly restricted by their memory cost, as we already used up to 500 million particles for simulations in the 4\,cm long gap.

In our simulations, we found that photoionization could sustain positive streamer propagation, but only in background electric fields close to the critical field.
Since there is considerably less photoionization in CO$_2$ than in air, the simulated discharges propagated more stochastically and they could eventually branch.
With a lower quenching pressure, these stochastic effects became stronger, the streamer velocity was lower, and the background field required for streamer propagation slightly increased.
In our experiments, streamers also developed much more stochastically in CO$_2$ than they did in air.
However, the observed velocities and diameters were about a factor of two higher than in the simulations, which we mostly attributed to the different electrode geometries that were used.

We have discussed the uncertainties in the CO$_2$ photoionization model, and we have presented an estimate for the quenching pressure based on the radiative lifetime of emitting states and the collision frequency of the gas.
From this estimate, we expect that the quenching pressure for photoionization in CO$_2$ is of comparable magnitude as the quenching pressure for photoionization in air.
However, an accurate estimate of this parameter will probably require new experiments on photoionization in CO$_2$ discharges.
Finally, a criterion for self-sustained discharge growth due to photoionization was proposed, based on the idea that a single photoionization event should produce on average at least one additional photoionization event.
This criterion agreed reasonably well with our computational results.









\section*{Acknowledgments}
The research was supported by the Fundamental Research Funds for the Central Universities of China (Grant No. xtr052023003), the State Key Laboratory of Electrical Insulation and Power Equipment of China (No. EIPE23114), and by the Dutch Research Council (NWO) through AES project 15052 `Let CO2 Spark' and AES project 20344 `Green Sparks'.

\section*{Data availability statement}
The data that support the findings of this study are available upon reasonable request.

\appendix
\section{Results from different runs}
\label{sec:diff_runs}

\begin{figure*}[htb]
	\centering
	\includegraphics[width=0.8\textwidth]{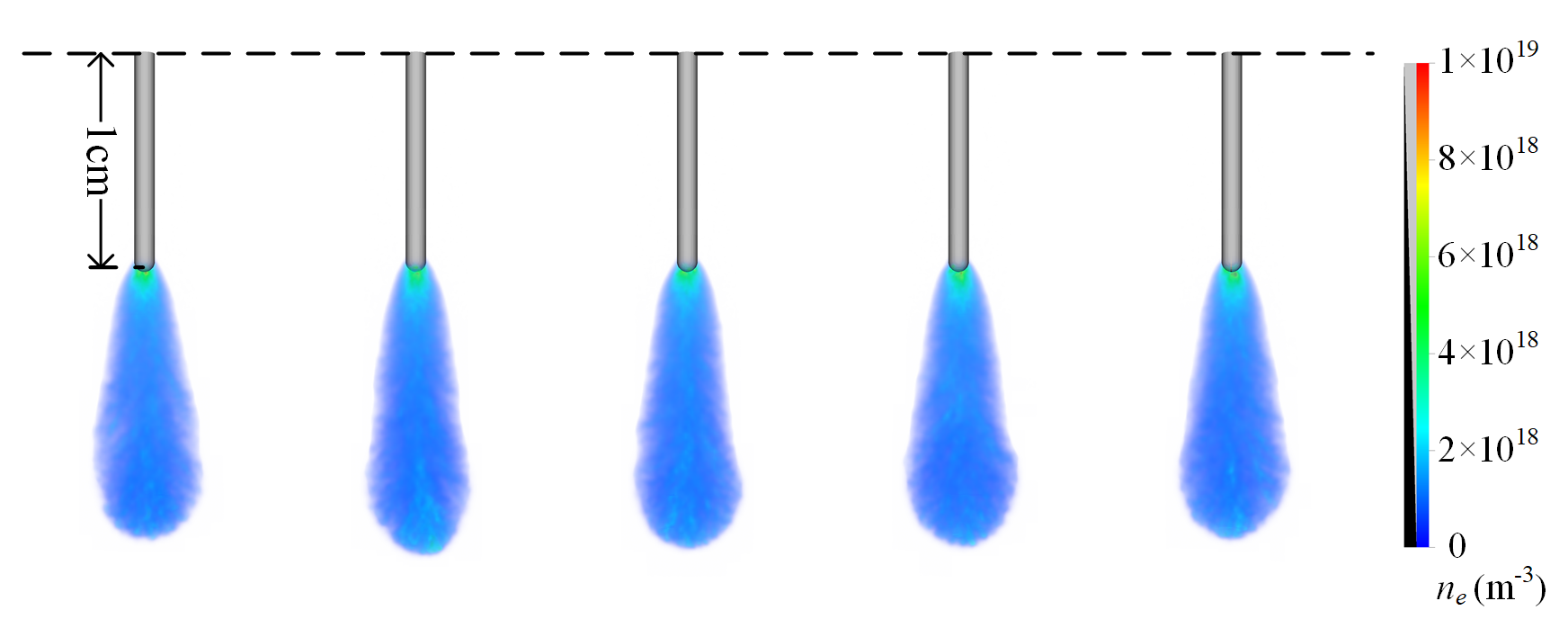}
	\caption{Electron density at 45\,ns of CO$_2$ streamers from different runs.}
	\label{fig:diff_runs}
\end{figure*}

To illustrate the run-to-run variation of the Monte Carlo model, we have simulated a CO$_2$ streamer five times using different random numbers.
The simulations were performed without quenching at 300\,K and 50\,mbar, using an applied voltage of 4\,kV.
As shown in figure~\ref{fig:diff_runs}, the streamer velocities are similar across multiple runs and this is also the case for the electron density in the streamer channel and the streamer radius.

\section*{References}

\bibliography{references_zotero, references_jannis}

\end{document}